\documentstyle[11pt,aaspp4]{article}

\lefthead{Wilkin}
\righthead{Non-Axisymmetric Bow Shocks}

\begin{document}

\title{Modeling Non-Axisymmetric Bow Shocks:\\
 Solution Method and Exact Analytic Solutions}

\author{Francis P. Wilkin}
\affil{California Institute of Technology, IPAC MS 100-22, Pasadena, CA 91125}
\affil{wilkin@ipac.caltech.edu}



\begin{abstract}
A new solution method is presented for steady-state, momentum-conserving, 
non-axisymmetric bow shocks and colliding winds 
in the thin-shell limit. This is a generalization of previous
formulations to include a density gradient in the pre-shock ambient
medium, as well as anisotropy  in the
pre-shock wind. For cases where the wind is unaccelerated, the formalism
yields exact, analytic solutions. 

Solutions are presented for two  bow shock cases: 
(1) that due to a star moving supersonically with 
respect to an ambient medium with a  density gradient perpendicular 
to the stellar velocity, and (2) that due to a star with a misaligned, 
axisymmetric wind moving in a uniform medium. It is also shown under
quite general circumstances that the total rate of energy thermalization
in the bow shock is independent of the details of the wind asymmetry, 
including the orientation of the non-axisymmetric
driving wind, provided the wind is non-accelerating and point-symmetric.
A typical 
feature of the solutions is that the region near the standoff point is tilted,
so that the star does not lie along the bisector of a parabolic fit 
to the standoff region. The principal use of this work is to infer
the origin of bow shock asymmetries, whether due to the wind or ambient
medium, or both.

\end{abstract}


\keywords{hydrodynamics --- ISM: bubbles --- ISM: HII regions---shock waves
---stars: mass loss}

%

\section{Introduction}

Supersonic stellar winds shock the surrounding gas and drive expanding bubbles
into the interstellar medium. These shocks provide an opportunity
to probe the properties of both the driving stellar wind and the
ambient medium. If the star is moving with respect to the interstellar gas,
the bubble will be distorted into a cometary shape. When the stellar
motion is supersonic, we refer to these as stellar wind bow shocks 
(Baranov, Krasnobaev \& Kulikovskii 1971; Dyson 1975). Since the discovery
of such bow shocks around young B stars (Van Buren \& McCray 1988), 
bow shocks have been found associated with many classes of objects,
such as pulsars (Kulkarni et al.~1992) and 
cataclysmic variables (e.g. Vela X-1: Kaper et al.~1997);
examples include well-known naked-eye stars
(e.g. Betelgeuse: Noriega-Crespo et al.~1997).
Bow shocks have been proposed as an explanation for cometary, ultracompact
HII regions (Van Buren et al.~1990; Mac Low et al.~1991) and as a means of 
explaining the lifetimes of ultracompact HII regions. 
In a recent survey of the IRAS database
using HiRes processing, Van Buren, Noriega-Crespo, \& Dgani (1995) 
found 58 candidate bow shocks.

Non-axisymmetric stellar wind bow shocks occur when a star with an anisotropic
wind moves supersonically with respect to the local medium, or if the star
has an isotropic wind but moves in an ambient medium containing a 
transverse density  gradient. 
Models of non-axisymmetric bow shocks are relevant to cometary ultracompact 
HII regions due to wind-blowing O stars moving supersonically with respect to 
the surrounding molecular cloud, when the ambient material does not have a 
constant density. 
A non-axisymmetric bow shock has also been invoked to explain
the morphology of Kepler's supernova remnant, where the supernova ejecta
collide with a non-axisymmetric bow shock generated by the pre-supernova wind
(Bandiera 1987; Borkowski, Blondin \& Sarazin  1992). Another example is  
the bow shock due to the head of a jet propagating 
into a region with a density gradient. 
Non-axisymmetric, ram-pressure balance models of the collision between a 
stellar wind and the photoevaporating flow from an externally 
illuminated circumstellar disk have been given by Henney et al.~(1996). 
A formulation for steady-state 
non-axisymmetric bow shocks and colliding winds was given by 
Bandiera (1993). However, Bandiera's numerical method is sufficiently
complicated that a simpler, analytic method is desirable. 

In this  contribution, I present a method for solving the problem 
of steady-state, momentum-conserving, non-axisymmetric,  thin-shell bow shocks 
and colliding winds. 
This is an extension of the previous analytic solution method of 
Wilkin (1996, hereafter Paper I) and  of 
Cant\'o, Raga, \& Wilkin (1996, hereafter Paper II) 
to non-axisymmetric  problems (see also Wilkin 1997a).

An outline of the paper is as follows. In \S~2, we formulate 
the problem of the steady-state collision of two winds,  and 
in \S~3 we treat the problem of a bow shock resulting from an isotropic wind 
interacting with a plane-parallel flow containing a transverse 
density gradient. In \S~4, we allow for non-isotropic winds, especially an 
axisymmetric wind with random orientation of the symmetry axis with
respect to the direction of stellar motion. The rate at
which kinetic energy is thermalized for the bow shock is treated
in \S~5. Results and future directions of this research are summarized
in \S~6.

\section{Mathematical Formulation and Solution Method}

\subsection{Description of the Collision Surface}

The hypersonic collision of two winds
will in general result in a system of two shocks.
A specific example, that of a stellar wind bow shock, is shown
schematically in Figure~\ref{fig:fig1}. 
Because this paper is concerned with steady-state solutions,
the colliding winds are assumed to be unchanging in time.
The stellar wind bow shock arises when a wind-blowing star moves
supersonically with respect to the intersellar gas. In this case,
we formulate the problem in the reference frame of the star, so the
ambient medium is described as a wind of parallel streamlines impinging
on the bow shock. For the collision of two winds in a binary star, we
will neglect orbital motion in order to consider a steady-state
problem in an inertial reference frame.
In steady-state, the amount of mass and momentum within a given volume
does not increase, and a flow pattern exists between the two shocks that
carries away the mass and momentum deposited by the colliding winds.
If the shocks are radiative, post-shock cooling lowers 
the temperature of the gas and leads to a large compression. 
We make the strong thin-shell assumption that cooling is so efficient 
that the shocked shell collapses to an infinitesimally thin layer,
with a finite surface density $\sigma$ of matter. For this thin-shell
assumption to apply, it is also necessary that the post-shock gas not be
supported by magnetic fields, which, if well-coupled to the gas,  
could maintain a finite thickness
even in the presence of efficient cooling. We assume there are no other
forces such as those due to radiation or magnetic fields.
The two shocked winds may in principle be separated by a contact 
discontinuity. However, there would then be supersonic shear across 
the interface that is expected to be unstable, leading
to a mixing of the two fluids. A detailed treatment of the mixing
is beyond the scope of this work. Instead, 
the mixing is assumed to be instantaneous, 
so the shell will have a unique velocity ${\bf V}_t$ at any
location within it. This velocity represents an average of the turbulent
fluctuations that would be present in a more detailed treatment. 
In steady-state, the geometric shape of the shell is unchanging in time, 
so the velocity of matter within the infinitely 
thin layer must be purely tangent to the
shell. There will, however, be acceleration normal to the shell, because
the fluid typically does not follow a straight path. 
The assumption that the incident streams are hypersonic, combined with 
the perfect cooling assumption, means that the flow within the shell 
will also be hypersonic, and pressure forces may be neglected in describing the
motions along the shell.\footnote{Clearly this assumption
must break down for real, finite temperature  systems 
near the stagnation point, because the tangential flow 
velocity in the shell vanishes at that point. However, pressure
forces depend upon the gradient of the pressure, which also vanishes
at the stagnation point, so the pressure forces may still
be small compared to momentum deposition in the stagnation
region (Wilkin 1997b).
In any case, beyond a small region near the stagnation point, 
the tangential flow in the shell will be supersonic, 
provided the incident flows are supersonic and the 
shocked fluid cools efficiently.} 
Defining a spherical coordinate system $(r,\theta,\phi)$, and denoting
the radius of the surface by $r = R(\theta,\phi)$,
a complete description of this idealized shell is given by specifying
the quantities $R, \sigma, {\bf V}_t$ as a function of position $(\theta,\phi)$
within the shell. There are in fact only four independent quantities, because
the condition that the motion within the shell be tangential implies
that one need only solve for two velocity components.

\subsection{Previous Treatment of the Problem}

The conservation laws of mass and  momentum may be used to derive 
the properties of the shell in terms of those of the two incident winds. 
Since  the momentum conservation law has three components, 
there will be four equations in four unknowns, or 
including the  condition of zero normal velocity, 
five equations in five unknowns. 
Because the assumption of vanishing thickness eliminates
one spatial variable, these equations will be partial differential 
equations (PDEs) in two spatial coordinates.  
Suitable boundary conditions must also be specified. 
In practice, this means identifying the stagnation 
(or standoff) point, where the two winds collide head-on, and where
for steady-state, the ram pressures of the two fluids balance. 
The equations may then be integrated away from the standoff point, 
following the motion of a fluid element. 
In order to begin the numerical integrations, one generally finds 
that an expansion about the conditions at the stagnation point is needed.

This approach was taken by Bandiera (1993),
who derived a set of PDEs in curvilinear
coordinates matched to the shape of the shell. These equations were
then solved under the assumption of radial, constant velocity, 
isotropic winds from two point sources. 
The moving stellar wind bow shock problem is formally obtained by taking the limit
of one source placed infinitely far away, while allowing its mass loss 
rate to be infinite so as to produce a finite density near the other star
(at the bow shock). 

Bandiera  noted that for the specific problems cited,
the motion within the shell would be along planes.
One may readily see this for the bow shock shown in Figure~1.
Consider a point on the bow shock surface, and the plane of constant
azimuthal angle containing it and the stellar trajectory. 
The radial wind striking the shell 
at this point has momentum lying in this plane. Similarly, the ambient 
medium striking this point has momentum lying in the plane. 
If the two incident streams mix instantaneously on impact, the resultant
momentum must also lie in the same plane. 
As the fluid flows along the shell, 
it continues to incorporate momentum contributions lying in the
same plane. 
The collision of radial winds emanating from two point
sources gives the same result. Thus, given the perfect mixing asumption, the
flow within the shell will lie along planes, 
a situation we refer to as {\it meridional flow}.
If the shocked fluid does not mix, the fluid
trajectories in the shell are not confined to a plane, and further work
is needed to solve this more complicated  
problem.\footnote{In the case of axisymmetric problems with shear, 
one may show that for divergent flow fields such as those encountered 
in bow shocks and colliding winds, the flow geometry and the fluxes 
of mass, momentum and angular momentum are the same as for the mixed case 
(Wilkin, Cant\'o and Raga 2000).} 

The ensuing geometric simplification allowed Bandiera to construct 
a two-dimensional grid, consisting of neighboring trajectories within 
the shell at a set of azimuthal angles. His computational method was to 
integrate the PDEs for mass, normal momentum, and tangential momentum, 
by marching along the trajectories within the shell 
as if the equations were ODEs.  By differencing neighboring solutions,
numerical values of the cross-streamline derivatives were
obtained and supplied  to the ODE integrator. Provided
the spacing of the two-dimensional grid is sufficiently fine, his method 
will yield accurate solutions. Bandiera postulated that
the dependence of the equations on derivatives of quantities across
streamlines was ``fictitious,'' although he did not succeed in eliminating 
it from the equations.

In this contribution,  the solution method will be simplified to 
an integral approach that avoids the need for PDEs, 
solving purely algebraic equations. 
Additionally, the solution may conveniently be obtained in terms of 
ordinary Cartesian, cylindrical polar, 
or spherical polar coordinates, and does not require a coordinate
system matching the shape of the shell.

\subsection{The Solution Method}

The solution method is based upon the observation   that 
thin shells driven by hypersonic winds are momentum-conserving 
in the vector sense (Paper I).  
In order to conserve momentum in steady-state, 
the momentum flux in the shell must be
the {\it vector sum} 
of the two incident momentum fluxes, integrated from the
standoff point to the point of interest. 
Such an integration is performed over the
area of the shell between two planes of constant azimuthal angle $\phi$, 
in the limit that their separation $\Delta \phi$ is infinitesimal (Fig.~1). 
For axisymmetric flow with no rotational
motion about the symmetry axis, no mass or momentum 
crosses these bounding planes, so the mass and momentum flowing within 
such a wedge must exactly equal the sum of the mass or momentum fluxes 
onto the external faces of the bounding surfaces (shocks).  
Given the known vector momentum flux, 
one may determine the shape of the shell, since the
fluid must move in the direction of its momentum.
While this description is complete,  the mathematics was subsequently 
simplified with the inclusion of an additional, angular momentum flux 
integral (Paper II). In Papers I and II, only constant velocity, 
isotropic winds were considered. 
For such winds, the vector momentum flux incident onto the
shocked shell is independent of the detailed shape of the shell, so
it is possible to specify the flux of momentum onto the shell from each 
side analytically. 
The methods of Papers I and II will now be extended to non-axisymmetric 
bow shocks and colliding winds, provided the flow is meridional.
For meridional flow, the bounding
surfaces are again planes of constant $\phi$, allowing us to
integrate the external fluxes and determine the internal fluxes within
the shell. For anisotropic, radial, constant velocity winds, 
such as those considered by Bandiera, 
the momentum flux will depend only upon the number
of streamlines intersected, and we will be able to
obtain solutions analytically.

Proceeding with the solution method, we note the
location of the stagnation point, where the two streams collide head-on.
For the two-wind collision, this point lies between the two stars, 
along the line connecting their centers. 
For the bow shock due to a wind-blowing star moving with respect
to the ambient medium, it will lie on the stellar trajectory,
in the direction of stellar motion. For either problem, the 
z-coordinate axis is chosen to contain  the stagnation point at $\theta =0$,
with the coordinate origin at the wind source (Fig.~1). In
what follows, the wind from the coordinate origin will be referred
to as ``the wind'', while the ``second wind'' may be either a radial 
wind from  a second point source, or 
the ambient medium in the frame of the moving star, 
where the ambient velocity is ${\bf V}_a = - V_a\,{\hat {\bf z}}$.
However, we will show explicit formulas only for the bow
shock, reserving detailed treatment of the binary star colliding
winds for a future paper. 
One should bear in mind that
the methods shown below will be applicable as well for that problem.
Returning to the description of a thin slice of the shell bounded
by planes of constant $\phi$,  the mass, momentum, and
angular momentum flux functions per unit azimuthal angle 
are defined by
$$ \Phi_m = \varpi\, \sigma \,V_t\, \sec\eta,\eqno(1)$$
$$ {\bf \Phi} = \Phi_m \,V_t \ {\hat {\bf t}},\eqno(2)$$
$$ {\bf \Phi}_J = {\bf R} \times {\bf \Phi}.\eqno(3)$$ 
Here $\varpi = R\sin\theta$ is the cylindrical radius, and
the unit vector at constant $\phi$ tangent to the
shell is  ${\hat {\bf t}}= {\hat {\bf \phi}}\times {\hat {\bf n}}/|
{\hat {\bf \phi}}\times {\hat {\bf n}}|$, where ${\hat {\bf n}}$
is the unit outward normal to the shell.
Noting that the arc length element $ds$ traced along the shell is given by
$(ds)^2 = (dR)^2+R^2(d\theta)^2+R^2\sin^2\theta\,(d\phi)^2,$
the arc length element traced along the shell at constant $\theta$
is $\varpi \sec\eta \,d\phi$, 
where the  angle $\eta$ is given by
$$ \tan\eta =  {{\partial R} \over {\partial \phi}}/R\sin\theta.\eqno(4)$$
For example, the flux of mass crossing a surface of constant polar 
angle $\theta$, between the azimuthal angles $\phi$ and $\phi+d\phi$,
is $\Phi_m d\phi$. 
If $V_\theta$ is the $\theta$-component of velocity 
within the shell,  and we write the cylindrical components of the
momentum flux function as $\Phi_\varpi$ and $\Phi_z$, then these
components are related to the angular momentum flux by
$$ {\bf \Phi}_J = \Phi_m\,R\, V_\theta\, {\hat {\bf \phi}},\eqno(5)$$
$$  {\bf \Phi}_J = R \,(\Phi_\varpi\cos\theta - \Phi_z \sin\theta)\, 
{\hat {\bf \phi}}.\eqno(6)$$ 
For meridional flow, the angular momentum within the shell 
will be in the ${\hat {\bf \phi}}$-direction,
so for the remainder of the discussion we will only
need the magnitude of the angular momentum flux $\Phi_J = 
{\bf \Phi}_J\cdot{\hat{\bf \phi}}$. 

The mass, momentum, and angular momentum 
flux functions due to the incident winds are similarly defined.
For the first wind, located at the coordinate origin, these are
given in terms of the wind density $\rho_w$ and velocity ${\bf V}_w$ as 
$$\Phi_{m,w} \Delta \phi = 
\int \int \rho_w ({\bf V}_w \cdot {\hat {\bf n}})\, dA,\eqno(7)$$
$${\bf \Phi}_{w} \Delta \phi = 
\int \int \rho_w ({\bf V}_w \cdot {\hat {\bf n}}) {\bf V}_w\, dA,\eqno(8)$$
$${\bf \Phi}_{J,w} \Delta \phi = 
\int \int \rho_w ({\bf V}_w \cdot {\hat {\bf n}})
({\bf R} \times {\bf V}_w)\, dA.\eqno(9)$$
The area of integration is that between two planes of constant $\phi$,
from the standoff point to polar angle $\theta$, following the shape
of the collision surface, which is determined along the course of the
integrations.  The sign of the normal direction is
chosen so as to point away from the origin.
The flux functions for the ambient medium are strictly analogous,
except the unit
vector  normal to the collision surface
would be in the reverse direction:
$$\Phi_{m,a} \Delta \phi = 
 - \int \int \rho_a ({\bf V}_a \cdot {\hat {\bf n}})\, dA,\eqno(10)$$
$${\bf \Phi}_{a} \Delta \phi = 
- \int \int \rho_a ({\bf V}_a \cdot {\hat {\bf n}}) {\bf V}_a\, dA,\eqno(11)$$
$${\bf \Phi}_{J,a} \Delta \phi = 
- \int \int \rho_a ({\bf V}_a \cdot {\hat {\bf n}})
({\bf R} \times {\bf V}_a)\, dA.\eqno(12)$$
If $R(\theta,\phi)$ is the radius of the shell, and its partial 
derivatives are $R_\theta$ and $R_\phi$, the surface
area element is
$$ dA = {\sqrt{R^2 + R_\theta^2
+ R_\phi^2\,\csc^2\theta}}\,
R\,\sin\theta\,d\theta\,d\phi.\eqno(13)$$
The unit outward normal ${\hat {\bf n}}$ to the surface 
is given by
$$ {\hat {\bf n}} = {{\bigl(R\,{\hat {\bf r}} 
- R_\theta\, {\hat {\bf \theta}}
-  R_\phi\,\csc\theta\, {\hat {\bf \phi}}\bigr)} \over {
{\sqrt{R^2 + R_\theta^2 + R_\phi^2\,\csc^2\theta}}}}.
\eqno(14)$$
Combining these, 
$$ {\hat {\bf n}}\, dA = \bigl(R\,{\hat {\bf r}} 
-R_\theta\, {\hat {\bf \theta}}
- R_\phi\,\csc\theta\, {\hat {\bf \phi}}\bigr)\,
R\,\sin\theta\,d\theta\,d\phi.\eqno(15)$$
For a radial wind, the partial derivatives with respect to $\theta$ and $\phi$ 
do not enter the expression for ${\bf V}_w \cdot {\hat {\bf n}}\,dA$, 
which is simply 
$$ {\bf V}_w \cdot {\hat  {\bf n}}\, dA = 
V_w\,R^2\sin\theta\,d\theta\,d\phi.\eqno(16)$$
Similarly, for the ambient medium ${\bf V}_a = -  V_a\, {\hat {\bf z}} = - V_a
({\hat {\bf r}}\cos\theta-{\hat {\bf \theta}}\sin\theta)$, so
noting that $d\varpi/d\theta=R\cos\theta+R_\theta\sin\theta$, we have
$$  - {\bf V}_a \cdot {\hat {\bf n}} \,dA = 
  V_a \,\varpi\, d\varpi\, d\phi.\eqno(17)$$
This eliminates some of the partial derivatives that complicated Bandiera's
very general treatment of the geometry. The rate at which 
conserved quantities 
are advected onto the shell does not depend upon the detailed description
of the shell, provided we ensure the correct number of streamlines
is counted.
Using eq.(16), the resulting forms of eqs.(7-9) 
for the radial stellar wind are
$$\Phi_{m,w}  =  \int_0^\theta 
R^2\, \rho_w\, V_w \, \sin\theta'\,d\theta',\eqno(18)$$
 $${\bf \Phi}_w     = 
\int_0^\theta R^2 \,\rho_w\, V_w^2 \,
[{\hat {\bf \varpi}}\,\sin\theta'   +    {\hat {\bf z}}\,\cos\theta'] \, 
\sin\theta' \, d\theta',\eqno(19)$$
$$ \Phi_{J,w} = 0.\eqno(20)$$
Here the radial unit vector has been written in terms of its 
cylindrical polar components, and 
the integrations are performed at constant $\phi$. 
Because the stellar wind is radial, it imparts
no angular momentum to the shell, and the wind angular
momentum flux vanishes. 
Using eq.(17) to simplify eqs.(10-12) for the ambient medium, 
$$\Phi_{m,a}  = 
  V_a\,\int_0^\varpi \rho_a  \varpi'd\varpi',\eqno(21)$$
$${\bf \Phi}_{a}  = 
-  V_a^2\,{\hat {\bf z}}\,\int_0^\varpi \rho_a \, \varpi'd\varpi',\eqno(22)$$
$${\bf \Phi}_{J,a}  = 
  V_a^2\,{\hat {\bf \phi}}\,\int_0^\varpi \rho_a \, {\varpi'}^2\,d\varpi'.\eqno(23)$$

The mass conservation law for the pre-shock radial wind in steady-state is
$${{\partial} \over {\partial r}} \biggl(r^2 \rho_w V_w\biggr) = 0.\eqno(24)$$
Consequently, if the wind is of mass loss rate ${\dot M}_w$ and streamline-average speed  ${\bar V}_w$, its properties are given by
$$ \rho_w V_w = {{{\dot M}_w} \over {4\pi\,r^2}} f_w(\theta,\phi),\eqno(25)$$
$$ \rho_w V_w^2 = 
{{{\dot M}_w\,{\bar V}_w} \over {4\pi\,r^2}} g_w(\theta,\phi).\eqno(26)$$
Here the dimensionless functions $f_w$ and $g_w$ are 
normalized to have unit streamline-average
over $4\pi$ steradians. 
Note that eq.(26) assumes the stellar wind to be coasting. 
Given this assumption, $g_w$ is independent of the radius. 
The wind mass and momentum fluxes onto the shell may now be written 
$$\Phi_{m,w}  =  {{{\dot M}_w} \over {4\pi}}\int_0^\theta 
f_w(\theta',\phi) \, \sin\theta'\,d\theta',\eqno(27)$$
 $${\bf \Phi}_w     = 
{{{\dot M}_w\,{\bar V}_w} \over {4\pi}}\int_0^\theta 
g_w(\theta',\phi)\,[{\hat {\bf \varpi}} \,\sin\theta' 
+   {\hat {\bf z}}\, \cos\theta'] \, 
\sin\theta' \, d\theta'.\eqno(28)$$

In order to obtain steady-state solutions, the ambient density
must be independent of the z-coordinate, although its dependence
on the remaining $\varpi,\phi$ coordinates may be arbitrary. Thus, we
assume an ambient density of the form
$$ \rho_a = \rho_0 \,f_a(\varpi,\phi),\eqno(29)$$
where $\rho_0$ is the value of the ambient density along the
stellar trajectory and $f(0,\phi)=1$. The final forms of the ambient
flux functions are thus
$$\Phi_{m,a}  =   \rho_0 \,V_a \, 
\int_0^{\varpi} f_a\, \varpi' \, d\varpi'.\eqno(30)$$
$${\bf \Phi}_a  = 
 - V_a\,{\hat {\bf z}}\, \Phi_{m,a}.\eqno(31)$$ 
$$ \Phi_{J,a} =  \rho_0\,V_a^2\, 
\int_0^\varpi f_a \, \varpi'^2 \, d\varpi'.\eqno(32)$$
The mass, momentum, and angular momentum  flux functions 
for both the wind and ambient medium are clearly streamline integrals that do 
not depend
upon the detailed shape of the shell - the integration is essentially
one over the solid angle of the $\Delta \phi$ wedge, as seen by the origin. 
This is a consequence of the fact that the pre-shock media conserve
these quantities in steady-state.  

Suppose we have performed the integrations  for the 
mass, momentum, and angular momentum onto the narrow slice of the shell,
for both incident winds, according to eqs.(20,27,28,30-32). In steady-state, 
these quantities do not accumulate at any location within
the shell, but are carried away by the flow within the layer. 
The conservation laws of mass, momentum, and angular momentum take the form
$$ \Phi_m =  \Phi_{m,w} + \Phi_{m,a},\eqno(33)$$
$$ {\bf \Phi} = {\bf \Phi}_w + {\bf \Phi}_{a},\eqno(34)$$
$$ \Phi_J = \Phi_{J,w} + \Phi_{J,a}.\eqno(35)$$
Now we may describe the properties of the shell explicitly in terms
of those of the external winds.  
Equation (6) yields the shell radius
$$ R = {{\Phi_J} \over {\Phi_\varpi\cos\theta-\Phi_z\sin\theta}}.\eqno(36)$$
This equation combines the formulation of Paper II, 
which was in terms of azimuthally integrated fluxes, 
with  the original treatment of Paper I, 
using fluxes in an infinitesimally thin wedge.
In eq.(36), each momentum flux function within the shell is to
be evaluated as the sum of appropriate source terms, using the
conservation laws (eqs.[34,35]). 
It is to be stressed that if the flow is not meridional, eq.(5)
does not hold, and the solution method is considerably more complicated. 
However, eqs.(6,36) are still valid provided the angular momentum
flux is replaced by its ${\hat {\bf \phi}}$ component. 
From eqs.(1,2), the tangential velocity of matter  in the shell is 
$$ V_t \,{\hat {\bf t}} = {\bf \Phi}/\Phi_m,\eqno(37)$$
while the mass surface density in the shell is given by
$$ \sigma = (\Phi_m^2/\varpi|{\bf \Phi}|) \cos\eta.\eqno(38)$$

\subsection{The Applied Torque Method and an Example}

To provide a specific example of the solution method, as well as a
reference solution to be compared to in the following, we consider
the simple problem of a bow shock from an isotropic wind in a
uniform ambient medium of density $\rho_a$. 
In this case, $f_w = g_w = f_a = 1$, and the flux functions are
$$\Phi_{m,w}  =  {{{\dot M}_w} \over {4\pi}} 
(1-\cos\theta),\eqno(39)$$
$${\bf \Phi}_w   = 
{{{\dot M}_w  V_w} \over {8\pi}}  \ 
[ {\hat {\bf \varpi}} \,(\theta - \sin\theta\cos\theta) 
+   {\hat {\bf z}}\, \sin^2\theta],\eqno(40)$$
$$ \Phi_{J,w} = 0,\eqno(20)$$
$$ \Phi_{m,a} = \frac{1}{2} \varpi^2 \,\rho_a\,V_a,\eqno(41)$$
$$ {\bf \Phi}_a = 
- \frac{1}{2} \varpi^2 \rho_a V_a^2 \,{\hat {\bf z}},\eqno(42),$$
$$\Phi_{J,a} = \frac{1}{3} \varpi^3 \rho_a V_a^2.\eqno(43)$$
Because the stellar wind has uniform speed, we have replaced  ${\bar V}_w$
with $V_w$. One could immediately substitute these results
into eq.(36) to determine the shell's shape.
However, before proceeding to apply the formalism, note that
the equation for the shape of the shell depends upon a specific 
combination of three flux functions in eq.(36).
One may divide the equation into two parts, one part for
each wind, according to
$${\cal T}_k = \varpi [\Phi_{\varpi,k} \cot\theta - \Phi_{z,k}] 
- \Phi_{{\rm J},k},\eqno(44)$$
and $k=w$ or $a$ for the wind and ambient (or second wind) sources. 
Physically, ${\cal T}$ represents the applied torque necessary
to compress the fan of streamlines between the stagnation point and
position $R(\theta,\phi)$ into a unidirectional 
beam possessing the same momentum flux (we shall call ${\cal T}$ the {\it
required torque}).
This is clearly seen by considering the radial wind from the coordinate
origin. 
It has no angular momentum, but an equivalent beam containing the
same momentum flux, located at a specific point $R(\theta,\phi)$ does
indeed have angular momentum about the origin, because the beam will not
be radial. The required torque 
for the isotropic wind  follows from 
eqs.~(20,40,44):
$$ {\cal T}_w =  - {{{\dot M}_w\, V_w \, \varpi} 
\over {8\pi}} (1- \theta \cot\theta).\eqno(45)$$
Using eqs.(41-44), 
the torque necessary to compress the ambient streamlines
to a unidirectional beam is
$$ {\cal T}_a = \frac{1}{6} \varpi^3 \rho_a V_a^2.\eqno(46)$$
For the ambient medium, the required torque expression simplifies because
the ambient medium has no cylindrical radial momentum, and its z-momentum
flux is related to its mass flux by eq.(31),
so in general,
$${\cal T}_a = \varpi\, V_a \,\Phi_{m,a} - \Phi_{J,a}.\eqno(47)$$
Similarly, the radial wind's required torque is simplified due
to its lack of angular momentum about the origin. 

The solution surface is now defined by 
$${\cal T}_w + {\cal T}_{a} = 0.\eqno(48)$$
The interpretation  of eq.(48) is that each wind supplies 
the torque necessary to compress the other wind. 
These torques are equal and opposite only if the correct
value of the radius (or cylindrical radius $\varpi$) is used, 
conserving angular momentum as well as linear momentum. 
The utility of this approach is
that if we vary the properties of one of the winds, holding the 
other constant, the shape of the shell is affected only by
the change in the applied torque from the second wind.

Now the applied torque expressions for the wind (eq.[45]) 
and the ambient medium (eq.[46]), 
when substituted into eq.(48), quickly recover the solution of Paper I,
$$ R(\theta) = R_0 \csc\theta {\sqrt {3(1-\theta \cot\theta)}},\eqno(49)$$
where $R_0$ is the standoff radius:
$$ R_0 = {\sqrt { {{{\dot M}_w V_w} \over {4 \pi \rho_a V_a^2}} }}.\eqno(50)$$

\section{Solutions for Bow Shock with Ambient Density Gradient}

In this section we assume the asymmetry of the bow shock to be due
to a density gradient in the ambient medium, while the  the stellar
wind is assumed to be isotropic. 
The next section will treat bow shocks from anisotropic winds.

The mass and momentum contributions of the stellar wind to the shell 
are given by eqs.(39,40,20), and its required torque is given by eq.(45).
The ambient medium is assumed 
to have a plane-parallel  density stratification, so in eq.(29), we
replace $f_a(\varpi,\phi)$ simply by $f_a(x)$, with 
$f_a$ is normalized so that $f_a(0) = 1$, and where
$$  x = \varpi \cos\phi.\eqno(51)$$
The integrated mass and angular momentum flux 
onto the wedge from the ambient medium are
$$\Phi_{m,a}  =   \rho_0 V_a 
\int_0^{\varpi}  f_a(\varpi' \, \cos\phi)\, \varpi'\,d\varpi'.\eqno(52)$$
$$ \Phi_{J,a} = \rho_0 V_a^2\, 
\int_0^\varpi f_a(\varpi'\,\cos\phi)\, \varpi'^2 \,d\varpi'.\eqno(53)$$
A complete description of the bow shock is now obtained by adding the
contributions of both the wind and ambient medium using the conservation
laws (eqs.[33-35]).

\subsection{Solution for Exponential Density Law}

We first obtain the solution for an exponential distribution
of ambient density. Let the density scale height be $H$.
Define for brevity the coordinate $y = x/H$, so the density law is 
$$ f_a(x) = \exp(-y),\eqno(54)$$
and the mass and angular momentum flux integrals are
$$\Phi_{m,a} =   H^2 \rho_0 V_a \sec^2\phi \,
[1 - \exp(-y)(1+y)],\eqno(55)$$
$$\Phi_{J,a} =   H^3 \rho_0 V_a^2 \sec^3\phi \,
[2 - \exp(-y)(y^2+2 y +2)].\eqno(56)$$
The ambient momentum flux follows from eq.(31), while the required
torque follows from eq.(47), which yields
$$ {\cal T}_a = H^3 \rho_0 V_a^2 \sec^3\phi 
\biggl[y - 2 + \exp(-y)(y+2)\biggr].\eqno(57)$$
Using this torque formula and that for the stellar wind given by eq.(45), 
the shell's shape is given by eq.(48), which yields upon simplification
$$ l^{2} \sec^2\phi \, y^{-1}[y-2 + \exp(-y) (y+2)] = {1 \over 2}
(1-\theta \cot\theta).\eqno(58)$$
Here $l = H/R_0$ is the density scale height in units of the standoff radius. 
Letting $y_{ax}$ be the value of $y = x/H$ appropriate for the
axisymmetric solution of eq.(49),
$$ y_{ax} = 
\frac{R_0}{H}\, \cos\phi\,{\sqrt{3(1 - \theta\,\cot\theta)}},\eqno(59)$$
eq.(58) takes the simple form
$$\frac{1}{y}\biggl[y - 2 + \exp(-y)(y+2)\biggr] = 
\frac{1}{6}\,y_{ax}^2.\eqno(60)$$
This formula may be solved numerically  for $y(\theta,\phi)$,
so letting $y_s$ denote the solution for $y$ to eq.(60), 
the shell's shape is given by
$$R(\theta,\phi) = H\, \sec\phi \csc\theta \ y_s(\theta,\phi).\eqno(61)$$ 
The result is a family of bow shock solutions distinguished 
by the  value of the nondimensional parameter $l$. 
By examining eq.(60) we may deduce several properties of the
solution. The left-hand side of the equation is monotonically
increasing as a function of $y$. The signs of $y$ and $y_{ax}$ must be
the same, since it is due to the $\cos\phi$ factor, 
so it follows that $y$ increases monotonically,
although in a very non-linear fashion, as a function of $y_{ax}$.
Because the parameter $l$ does not enter eq.(60), we see that there
is one universal solution for the problem, although this solution
is scaled by the value of $R_0$ and stretched (distorted)  
depending upon the value of $H/R_0$. 
The solution for the radius at all angles $\theta,\phi$ and for all
values of $l$ follows from Fig.~\ref{fig:fig2}. In particular, as $y\rightarrow
\infty$, the left hand side of eq.(60) approaches unity, implying
a breakout angle for the shell when $y_{ax}=\sqrt{6}$, which yields
$$ 1 - \theta_\infty \cot\theta_\infty  = 2\, l^2\, \sec^2\phi.\eqno(62)$$
The opening angle $\theta_\infty$
depends upon $\cos\phi$, and applies only to $\cos\phi \ge 0$. 
In the tail of the bow shock, for $\cos\phi < 0$, we have
$y \rightarrow - \infty$. In this case, $y$ depends logarithmically
upon  $y_{ax}$. The most interesting part of the solution, for $|y| \leq 1$
may be described by an expansion in terms of small $y_{ax}$.
Noting that
for $\cos\phi=0$ the solution is identical to that of eq.(49),
we anticipate that an expansion for small $\cos\phi$ corresponds
to an expansion in terms of small $y_{ax}$
about the standard bow shock solution.
Letting $R_{ax}$ be the value of $R(\theta)$
for that solution (given by eq.[49]), we obtain
$$R \approx R_{ax}\biggl[1 + \frac{1}{4}\,y_{ax}
+ \frac{13}{160}\,y_{ax}^2+\frac{7}{240}\,y_{ax}^3
+\frac{11843}{1075000}\,y_{ax}^4 +....\biggr].\eqno(63)$$
The behavior of the solution near the stagnation point is given by
$$  R = R_0 \,
\biggl[1 +  {{\cos\phi} \over {4\,l}} \theta + 
\bigl({1 \over 5} + {{13 \cos^2\phi} \over {160\,l^2}}\bigr) \theta^2 
 + ...\biggr].\eqno(64)$$
Unlike the axisymmetric bow shock, there is a linear term
in the behavior near the standoff point, so it is not describable as 
parabolic with the z coordinate axis as the axis of a parabola, except for the
special angles $\phi = \pi/2 \ {\rm or}\ 3\pi/2$. 
As a consistency check, note that as the scale height increases relative
to the standoff radius $(l=H/R_0\rightarrow \infty)$, the solution reduces
to the standard axisymmetric bow shock (Paper I). 
The lowest order effect near the standoff point is a 
{\it tilt} of the bow shock head, described by
the linear term in $\theta$. 
This means that although the wind and ambient streamlines
meet head-on at the standoff point, they are not normal to the
shell at this location, quite different from the axisymmetric case.
This effect is due to the instantaneous mixing we have assumed. If shear
is present in the shell, the two colliding flows could
have separate stagnation points, which would be located where the incident
stream is indeed normal to the shell. 
These bow shock solutions differ from the standard model (Baranov, Krasnobaev
\& Kulikovskii 1971) in that the exponential ambient density distribution
implies a finite total mass on one side of the bow shock. The solution then
more closely resembles those of two-wind collisions (Paper II) 
where there is a  finite opening angle for the bow shock tail. 
Similar problems include blast
waves in finite mass media  (Koo \& McKee 1990) and wind breakout from 
a stratified medium (Cant\'o 1980; Borkowski, Blondin \& Harrington 1997). 
For the high density region, $\cos\phi < 0$, the shell's asymptotic 
shape in the tail region is
$$ z \approx - \frac{2 l^2}{\pi}\varpi\sec^2\phi 
\,\exp\bigl(-\frac{\varpi}{H}\cos\phi\bigr).\eqno(65)$$
The shell continues to be more distorted as it expands due
to the $\exp(-y)$ factor. 
Because the shell has a finite opening angle on one side, not all wind
streamlines intersect the shell, but those with $\theta>\theta_\infty(\phi)$
freely expand to infinity.  Further consequences of the finite opening
angle of the shell are discussed in \S 5.

\subsection{Solution for Linear or Polynomial Density Law}

Consider a stratification of the ambient medium according to
$$ f_a(x) = 1 + a_1 x + a_2 x^2 + ....\eqno(66)$$
Of course, one must ensure that this expression is non-negative over
the domain of interest. 
The fluxes of mass and angular momentum onto the shell are
$$\Phi_{m,a} =  \varpi^2 \rho_0 V_a  \, 
\biggl[{1 \over 2}  + {{a_1} \over 3} x + {{a_2} \over 4} x^2 + ...\biggr],\eqno(67)$$
$$\Phi_{J,a} =  \varpi^3 \rho_0 V_a^2  \,
\biggl[{1 \over 3} + {{a_1} \over 4} x + {{a_2} \over 5} x^2 + ...\biggr].\eqno(68)$$
The ambient momentum flux follows from eq.(31), while the required torque
follows from eq.(47), giving the result
$$ {\cal T}_a = {{\varpi^3} \over 6} \rho_0 V_a^2 
\biggl[1 + {{a_1} \over 2}  x + {3 \over {10}} a_2 x^2 + ... 
+ {{6 (n+1)!} \over {(n+3)!}} a_n x^n 
+ ... \biggr].\eqno(69)$$
Substitution into eq.(48), we obtain upon  simplification
$$ {{\varpi^2} \over {R_0^2}} 
\biggl[1 + {{a_1} \over 2} \varpi \cos\phi + {{3\,a_2}
\over {10}}\,\varpi^2\cos^2\phi + ...\biggr]
 = 3 \, (1 - \theta \cot\theta).\eqno(70)$$
Restricting the treatment to a linear density gradient, with only $a_1$ 
non-zero, this  equation is  cubic  in the variable $\varpi$.
One may  solve analytically or numerically for the function
$\varpi(\theta,\phi)$, so the solution surface is then given
by $R(\theta,\phi) = \varpi(\theta, \phi) \csc\theta$.
Inclusion of higher order terms in the density law simply increases
the degree of the polynomial, but the solution is obtained in the same way.

The behavior of the solution near the stagnation point is given by
$$ {R \over {R_0}} =  
1  - {{a_1 R_0 \cos\phi} \over 4} \theta + 
\biggl({1 \over 5} + ({5 \over {32}}
a_1^2 - {3 \over {20}} a_2) R_0^2\cos^2\phi\biggr) \theta^2 
+ ....\eqno(71)$$
As a consistency check of the solutions, 
for the case of a uniform ambient medium $(a_1,a_2,... =0)$, 
eqs.(70,71) reduce to the standard solution of Paper I.
As a further check, note that this result is consistent with the solution
for an exponential mass distribution, if we choose the coefficients 
corresponding to the exponential distribution of the previous subsection,
$a_1=-1/H,a_2=1/2 H^2,...$, recovering eq.(64).

\section{Treatment of Anisotropic Winds}

\subsection{The Axisymmetric Wind}

We now relax the assumption that the wind is isotropic to permit an
axisymmetric wind, where the axis of symmetry is misaligned with the
direction of stellar motion, or in the case of a two-wind collision
in a binary star, where the axis does not point to the other star.
Treatment of the accelerated wind will be deferred to a future contribution;
in this paper the wind is assumed to be coasting.
The coordinate axes are chosen so that the z-direction points
in the direction of stellar motion, or towards the second star. 
In terms of a spherical coordinate system, where the azimuthal angle
$\phi$ is measured about the z-axis, motion of the shocked fluid in the
shell is along planes of constant $\phi$. 
We also define starred coordinates so that the $z_*$-axis is the symmetry
axis of the stellar wind.  The two coordinate systems are related by
\setcounter{equation}{71}
\begin{eqnarray}
 \sin\theta_*\cos\phi_* & = & \sin\theta\cos\phi,\\
 \sin\theta_*\sin\phi_* & = & \sin\theta\sin\phi\cos\lambda 
 -  \cos\theta\sin\lambda, \\
 \cos\theta_* & = & \sin\theta\sin\phi\sin\lambda 
 +  \cos\theta\cos\lambda.
\end{eqnarray}

The wind mass and momentum flux densities depend on the
polar angle $\theta_*$:
$$ \rho_w V_w = {{{\dot M}_w} \over {4 \pi\, r^2}}\,f_w(\theta_*),\eqno(75)$$
$$ \rho_w V_w^2 = 
{{{\dot M}_w\,{\bar V}_w} \over {4 \pi \,r^2}} \,g_w(\theta_*).\eqno(76)$$
The  nondimensional functions $f_w$ and $g_w$ are normalized to have 
unit average value  over $4\pi$ steradians.

The incident fluxes of mass, momentum, and angular
momentum from the wind onto  the shell are
\setcounter{equation}{76}
\begin{eqnarray}
\Phi_{m,w}(\theta,\phi) & = & 
{{{\dot M}_w} \over {4 \pi}}   
F_w(\theta,\phi), \\
{\bf \Phi}_w(\theta,\phi) & = & 
{{{\dot M}_w\,{\bar V}_w} \over {4 \pi}} 
\,{\bf G}_w(\theta,\phi), \\
\setcounter{equation}{20}
\Phi_{{\rm J},w}(\theta,\phi) &  = & 0.
\end{eqnarray}
where the nondimensional functions $F_w$ and ${\bf G}_w = 
G_{w,\varpi}\,{\hat {\bf \varpi}} + G_{w,z}\,{\hat {\bf z}}$
are given by
\setcounter{equation}{78}
\begin{eqnarray}
F_w & = & 
\int_0^\theta f_w
\, \sin\theta' {\rm d}\theta', \\
{\bf G}_w & = & 
\int_0^\theta g_w
\,[{\hat {\bf \varpi}}\, \sin\theta' 
+  {\hat {\bf z}}\, \cos\theta']\,  
\sin\theta' {\rm d}\theta'.
\end{eqnarray}
Here  $f_w$ and $g_w$ have argument $\theta_*(\theta',\phi)$,
so the integrations are performed using the transformation eqs.(72-74) 
to evaluate $\theta_*(\theta,\phi)$.
We also define a nondimensional function $T_w$ associated with the
required torque ${\cal T}_w$ to compress the wind streamlines 
according to
$$ {\cal T}_w = 
{{{\dot M}_w\,{\bar V}_w} \over {4 \pi}} \varpi\,T_w.\eqno(81)$$
Because the wind's angular momentum flux vanishes, eq.(44) implies
that
$$ T_w = G_{w,\varpi} \cot\theta - G_{w,z}.\eqno(82)$$

The functions $f_w(\theta_*),\,g_w(\theta_*)$ may now be expanded
in terms of powers of
$\cos\theta_*$,
$$ f_w(\theta_*) = \sum_{i=0}^\infty b_i \cos^i\theta_*,\eqno(83)$$
$$g_w(\theta_*) = \sum_{i=0}^\infty c_i \cos^i\theta_*.\eqno(84)$$
Defining for brevity
$$ p = \sin\phi\sin\lambda,\eqno(85)$$
$$ q = \cos\lambda,\eqno(86)$$
the transformation given by eq.(74) 
is $\cos\theta_* = p \sin\theta + q \cos\theta$. 
Denoting the trigonometric integrals by
$$ {\rm I}_{j,k} = \int_0^\theta 
\sin^j\theta'\cos^k\theta' d\theta',\eqno(87)$$
we evaluate $F_w$ and ${\bf G}_w$  using the above series, and
bringing $p$ and $q$ outside of the integrals, we finally have
$$ F_w   = 
\sum_{i=0}^\infty b_i \sum_{j=0}^i
p^{i-j}\,q^j {i\choose j}\,I_{1+i-j,j},\eqno(88)$$
$$ {\bf G}_w   = 
\sum_{i=0}^\infty c_i \sum_{j=0}^i
p^{i-j}\,q\,^j {i\choose j}
\bigl[{\hat {\bf \varpi}}\,I_{2+i-j,j} + {\hat {\bf z}}\,I_{1+i-j,1+j}\bigr]
.\eqno(89)$$
Here we have used the binomial coefficients to write the sums.
The  applied torque necessary to compress the wind streamlines
to a thin shell, in nondimensional form, is now
$$  T_w = \sum_{i=0}^\infty c_i \, T_w^{(i)},\eqno(90)$$
where the responses due to the individual $cos^i\theta_*$ terms
are given by
$$T_w^{(i)} =  \sum_{j=0}^{i} p^{i-j} q^j {i\choose j}
\bigl[\cot\theta \, 
{\rm I}_{2+i-j,j} - {\rm I}_{1+i-j,1+j}\bigr].\eqno(91)$$

\subsection{General Solution for Bow Shock Driven by an Axisymmetric,
Misaligned Wind}

The wind description of the previous subsection may now 
be applied to the problem of a bow shock driven by an anisotropic
wind. The wind is assumed to be 
driven by a star moving at speed $V_a$ in
a medium of uniform density $\rho_a$. We describe the bow shock's 
properties in the frame of the star.
 
The standoff radius, defined as the shell radius at $\theta=0,$ which
corresponds to $\theta_*=\lambda$, is given by
$$ R_\lambda = 
{\sqrt{ {{{\dot M}_w {\bar V}_w g_w(\lambda)}\over 
{4 \pi \rho_a V_a^2}}}}.\eqno(92)$$
We also define $R_0$ to be the standoff radius for the equivalent
isotropic wind
$$ R_0 = 
{\sqrt{ {{{\dot M}_w {\bar V}_w}\over 
{4 \pi \rho_a V_a^2}}}}.\eqno(93)$$
It is important to recall for the results below 
that the standoff radius will be
$R_0 \sqrt{g_w(\lambda)}$, where
$$ g_w(\lambda) = \sum_{i=0}^\infty c_i\,\cos^i\lambda.\eqno(94)$$
By eqs.(46,48,82),
the solution for the shell's radius is 
$$ R(\theta,\phi) 
=  R_0 \csc\theta 
{\sqrt{- 6\, T_w}},\eqno(95)$$
where $T_w$ is given by eqs.(90,91).
The total mass and momentum fluxes, 
including the contribution from the ambient medium, are
$$ {\Phi}_m = {{{\dot M}_w} \over {4\pi}} 
\bigl[F_w+\frac{1}{2\,\alpha}{\tilde \varpi}^2\bigr],\eqno(96)$$
$$ {\bf \Phi} = 
\frac{{\dot M}_w {\bar V}_w}{4\pi}\bigl[G_{w,\varpi} \,{\hat {\bf \varpi}}
+ (G_{w,z} - \frac{1}{2}{\tilde \varpi}^2)\, {\hat {\bf z}}\bigr],\eqno(97)$$
where ${\tilde \varpi} = \varpi/R_0$, and 
$\alpha = V_a/{\bar V}_w$.

To obtain a complete description of the bow shock's properties, we 
must specify the velocity of material in the shell and the mass surface
density of matter. The velocity is given by the ratio of the
mass and momentum fluxes:
$$  V_t \,{\hat {\bf t}} = 
V_a {{\bigl[2\,G_{w,\varpi} \,{\hat {\bf \varpi}}
+ (2\,G_{w,z} - {\tilde \varpi}^2)\, {\hat {\bf z}}\bigr]} \over
{\bigl[2\,\alpha\,F_w+{\tilde \varpi}^2\bigr]}}.\eqno(98)$$
The surface density is given by eq.(38), which yields
$$ \sigma = \rho_a R_0  
{{\bigl[2\,\alpha F_w + {\tilde \varpi}^2\bigr]^2}\over
{2\,{\tilde \varpi}{\sqrt{\bigl[4\,G_{w,\varpi}^2+(2\,G_{w,z}-{\tilde \varpi}^2)^2
\bigr]}}}}\cos\eta.\eqno(99)$$
In order to evaluate $\sigma$, we need to know the angle $\eta$.
Using eqs.(4,95), we have
$$ \tan\eta = \frac{1}{2}\csc\theta \frac{1}{T_w}\frac{\partial\,T_w}{\partial\,\phi},\eqno(100)$$
where differentiation of eqs.(90,91) with respect to $\phi$, which
enters only in the variable $p$, yields
$$ \frac{\partial\,T_w}{\partial\,\phi}  = \sum_{i=1}^\infty
c_i \, {{\partial T_w^{(i)}} \over {\partial \phi}},\eqno(101)$$
$${{\partial T_w^{(i)}} \over {\partial \phi}} = 
\cot\phi
\sum_{j=0}^{i-1}(i-j) p^{i-j} q^j {i\choose j}
\bigl[\cot\theta \, 
{\rm I}_{2+i-j,j} - {\rm I}_{1+i-j,1+j}\bigr].\eqno(102)$$
Note that the summation limits are restricted so that the terms
with $j=i$ vanish, including the $c_0$ term.

In the region of the standoff point, an expansion 
to second order in $\theta$ gives
$$ R \approx  R_\lambda\,
\biggl\{1 + \theta \,\frac{p\,g_w'(\lambda)}{4\,g_w(\lambda)}
+ \theta^2\,\biggl[\frac{1}{5} 
 - \frac{p^2\,g_w'^2(\lambda)}{32\,g_w^2(\lambda)}$$
$$+ \frac{3\,(p^2\,g_w''(\lambda)
-q\,g_w'(\lambda))}{40\, g_w(\lambda)}\biggr]
\biggr\}.\eqno(103)$$
Primes on the function $g_w$ represent differentiation
with respect to $q = \cos\lambda$.
As was the case of asymmetry resulting from an ambient density
gradient,  there is a $\theta^1$ term indicating a tilt
of the standoff region relative to the direction of the stellar motion.
The mass and momentum flux functions near the standoff point are
$$ \Phi_m \approx {{{\dot M}_w} \over {4 \pi}} 
\biggl\{\frac{1}{2}\,f_w(\lambda)
+ \frac{1}{2\alpha}\,g_w(\lambda)\biggr\}\,\theta^2.\eqno(104)$$
$$ {\bf \Phi} \approx {{{\dot M}_w\,{\bar V}_w} \over {4 \pi}} 
\biggl\{\frac{1}{3}\,g_w(\lambda){\hat {\bf \varpi}} 
+ \frac{p}{3}\,g_w'(\lambda)\,{\hat {\bf z}}\biggr\}\,\theta^3.\eqno(105)$$
Equations~(4,28,29) now give the tangential velocity in the shell
and the mass per unit area as
$$ V_t \,{\hat {\bf t}} \approx \frac{2}{3}\, V_a \,\theta \,
\frac{ g_w(\lambda)\,{\hat {\bf \varpi}}
+ p\,g_w'(\lambda)\,{\hat {\bf z}}}{\alpha\,f_w(\lambda)+g_w(\lambda)},\eqno(106)$$
and
$$ \sigma  \approx \frac{3\,R_0\,\rho_a}{4}\,
\frac{(\alpha\,f_w(\lambda)+g_w(\lambda))^2}{{\sqrt{g_w(\lambda)\,(g_w^2(\lambda)+p^2\,(g_w'(\lambda))^2}}}.
\eqno(107)$$
One may readily see that these solutions reduce to the results
of Paper I for an isotropic wind ($f_w = g_w = 1$). 

\subsection{Simple Solutions: Quadratic Dependence on $cos\theta_*$}

The simplest non-trivial solution is for a linear
dependence of mass and momentum fluxes upon $\cos\theta_*$.
However, in astrophysical applications one is frequently
concerned with stellar winds that have symmetry with respect
to the equatorial plane, such as when the asymmetry
arises due to rotation of the star. For winds that are symmetric with 
respect to $\theta=\pi/2$, the functions $f_w(\theta_*),\,g_w(\theta_*)$ 
will have  expansions in terms of even powers of $\cos\theta_*$ only.
Thus, we shall give the solution for a wind that depends upon $\cos^2\theta_*$.
This is sufficiently general to include as a subset previous models
of non-axisymmetric bow shocks (Bandiera 1993; Chen \& Huang 1997).
We assume the mass and momentum fluxes are described
by 
$$f_w = b_0  + b_1 \cos\theta_* + b_2 \cos^2\theta_*,\eqno(108)$$ 
$$g_w = c_0  + c_1 \cos\theta_* + c_2 \cos^2\theta_*,\eqno(109)$$ 
where the normalization requires $b_0=(1-b_2/3)$ and $c_0 = (1-c_2/3)$.

The mass flux integral $F_w$ is 
$$ F_w = 
\biggl\{b_0'\,(1-\mu)
+ \frac{b_1}{2}[p\,(\theta-\sin\theta\cos\theta)+q \sin^2\theta]$$
$$ + \frac{b_2}{3}\sin^2\theta\,[(q^2-p^2)\cos\theta
+ 2 p q  \sin\theta]\biggr\},\eqno(110)$$
where $b_0' = b_0 + b_2 (2p^2+q^2)/3$.
The components of ${\bf G}_w$ are
$$G_{w,\varpi} = \frac{1}{4}\biggl\{c_0' (2\theta - \sin2\theta)
+ \frac{c_1}{3} \biggl[p\,(8-9\cos\theta+\cos3\theta)
+ 4 q \sin^3\theta\biggr]$$
$$ + c_2 \sin^3\theta \,\biggl[2 p q \sin\theta 
+ (q^2-p^2)\cos\theta\biggr]\biggr\},\eqno(111)$$
$$G_{w,z} = \frac{1}{4}\biggl\{c_0'(1-\cos2\theta) 
+ \frac{c_1}{3} \biggl[4 p \sin^3\theta
+ q\, (4-3\cos\theta - \cos3\theta)\biggr]$$
$$ + c_2\biggl[\frac{p q}{4}(4\theta-\sin4\theta)
+ \frac{(q^2-p^2)}{2}
(2+\cos2\theta)\sin^2\theta\biggr]\biggr\}.\eqno(112)$$
Here the coefficient $c_0' = c_0 + c_2 (3 p^2 + q^2)/4$ depends
on $\phi$ and $\lambda$.
Using eqs.(82,111,112) we obtain
$$ T_w = \biggl\{-\frac{c_0'}{2}(1-\theta\cot\theta) 
- \frac{c_1}{3}(1-\cos\theta)[q+p\tan(\frac{\theta}{2})]$$
$$ + \frac{c_2}{8}
\biggl[(p^2-q^2)\sin^2\theta - p q (2\theta - \sin2\theta)
\biggr]\biggr\}.\eqno(113)$$
Using eq.(95), with $T_w$ given by eq.(113), we obtain the shell's shape
$$R = R_0 \csc\theta  
\biggl\{3 (1 - \theta \cot\theta)(c_0 + \frac{c_2}{4} (3 p^2 + q^2))$$
$$+ 2 c_1 (1 - \cos\theta) \bigl[q + p \tan\bigl(\frac{\theta}{2}\bigl)\bigl]$$
$$ + \frac{3\, c_2}{4} 
\bigl[(q^2-p^2) \sin^2\theta+  p q (2\theta -\sin2\theta)\bigr]\biggr\}^{1/2},\eqno(114)$$
where the explicit dependence upon stellar inclination
and azimuthal angle is obtained using $p=\sin\phi\sin\lambda$ and 
$q=\cos\lambda$.  Examples are shown in Figures 4-6 
for even parity winds with $c_2 = 3, 1.5$, and $-1.5$.
Noting that Bandiera's notation for this
problem was $\Delta_Y = - 2 c_2/3$, these correspond to his cases of 
$\Delta_Y = -2,-1$ and $+1$.

In the neighborhood of the standoff point, the shell's shape is given by
$$ R^2/R_0^2 \approx \bigl[(c_0 + c_1 q + c_2 q^2) + 
\frac{p}{2}(c_1 + 2 c_2 q) \theta$$
$$+ (8 c_0 + 5 c_1 q + 2 c_2 (3 p^2 + q^2)) \theta^2/20\bigr].\eqno(115)$$

\subsection{The Non-Axisymmetric Wind}

The previous formulation is not restricted to an axisymmetric wind, so one
may obtain solutions for bow shocks driven by radial, non-axisymmetric winds,
by defining the momentum flux from the wind in terms of spherical harmonics.
Because the method is clear from the previous discussion of an 
axisymmetric wind, we will not give further details here. We note only
that the response from each spherical harmonic contribution to
the wind momentum flux may be  
obtained from the required torque formula. Then the shape of the 
shell will depend upon the square root of the sum of these responses.

\section{Energy Thermalization Rates}

The energy thermalization rate of axisymmetric bow shocks
and colliding winds has been considered by Wilkin, Cant\'o
\& Raga (2000). The thermalization occurs in the shocks,
post-shock (radiative) relaxation, and mixing. The maximum
possible rate of thermalization is given by the total incident
kinetic energy flux in the center of mass frame. Here we are concerned
with the total rate of energy thermalized, over the entire shell,
rather than the amount per unit area. By center of mass, we
mean the center of mass of the matter deposited onto the shell per
unit time. This maximum
thermalization rate is for the case of complete mixing. For the
bow shock in a uniform ambient medium, 
the center of mass frame corresponds to the ambient frame, 
because a formally infinite amount of ambient mass per unit time 
strikes the shell, while the stellar wind has finite mass loss rate. 
For the bow shock driven by a star with a radial wind,
the total energy thermalization rate is given by
$$ {\dot E}_{\rm therm} =  \int_0^{2\pi} \int_0^\pi \frac{1}{2} 
R^2 \rho_w \, V'^2_{w}\, 
({\bf V}'_w - {\bf V}'_{\rm shell}) \cdot {\hat {\bf r}}\,
\sin\theta\, d\theta\, d\phi.\eqno(116)$$
Here primes indicate velocities in the ambient frame, so ${\bf V}_w'
= V_w\, {\hat {\bf r}} + V_a\, {\hat {\bf z}}$, giving
$$ V'^2_{w} = [V_{w}^2 
+ 2 V_{w} V_a \cos\theta + V_*^2].\eqno(117)$$
The velocity of the shell is its pattern speed 
and corresponds to the stellar velocity ${\bf V}_* = V_a\,{\hat {\bf z}}$.
Because we are assuming a non-accelerating wind, the
integral depends only upon the streamlines and not the detailed shape
of the shell, so we may perform the integration over a spherical surface.

For the case of an isotropic wind from a star moving through 
a plane-parallel stratified ambient medium, the rate of energy 
thermalization is precisely the same as for a uniform medium 
(Wilkin, Cant\'o, \& Raga 1997),
since it is the kinetic energy in the 
ambient frame that is thermalized:
$$ {\dot E}_{\rm therm} = \frac{1}{2} {\dot M}_w (V_w^2 
+ V_*^2).\eqno(118)$$
This rate of energy thermalization is appropriate for an arbitrary
ambient distribution of matter, provided only that it is truly infinite
in mass, and that the bow shock is steady-state in the star's frame. 
The second requirement demands that the  ambient mass distribution be
independent of the z-coordinate, although its $(\varpi,\phi)$ 
distribution is arbitrary.
An exception, for example, is the case of an exponential 
stratification, for which case the total ambient mass flux may be finite
(on one side of the bow shock). For this case, the center of mass frame
is not equal to the ambient frame, and it is important to separately define
the center of mass frame for each $\phi$ slice, as it will 
depend upon azimuthal angle for the portion of the bow shock that
contains finite opening angle.
Because of the resulting opening angle, not all of the stellar
wind kinetic energy (in the ambient frame) will be thermalized, for two
reasons: (1) some streamlines miss the shell, and (2) 
there is a net flow of momentum along the shell even at infinite distance,
whereas for the standard bow shock $V_t \rightarrow 0$ in the ambient
frame, far in the bow shock tail.

We now consider the case of an anisotropic wind, to determine whether 
the energy thermalization rate depends on the orientation of the wind.
However, the thermalization rate remains independent of the stratification 
of the ambient medium, subject to the caveat about finite-mass systems.
For the remainder of this section, we allow the wind to be non-axisymmetric,
so in equations (75,76) we use 
$f_w = f_w(\theta_*,\phi_*),\,g_w=g_w(\theta_*,\phi_*)$. When we refer
to an axisymmetric wind, its axis of symmetry will be $\theta_*=0$ as 
before.

The total kinetic energy loss rate of the wind is 
$$ {\dot E}_w  =  \frac{1}{8\pi} {\dot M}_w\, {\bar V}_w^2
\int_0^{2\pi}\int_0^\pi {{g_w^2} \over {f_w
}} \sin\theta_*\, d\theta_*\,d\phi_*.\eqno(119)$$
The functions $f_w$ and $g_w$ are assumed to have unit average over
$4\pi$ steradians.
In terms of this energy loss rate, 
we define the mean square wind speed as 
$$<V_w^2>\, =\, 2 {\dot E}_w/{\dot M}_w.\eqno(120)$$
We also define a {\it total} vector momentum flux for the wind, a quantity
that is non-vanishing only if the wind is not point-symmetric with
respect to the origin
$$ {\bf P}_w = \frac{{\dot M}_w {\bar V}_w}{4\pi}\int_0^{2\pi}\int_0^{\pi} 
g_w \, {\hat r} \sin\theta d\theta\,d\phi,\eqno(121) $$
$$= \int_0^{2\pi} {\bf \Phi}_w(\pi,\phi) d\phi.\eqno(122)$$

We now wish to calculate the kinetic energy deposition rate to the shell
in the ambient frame. Letting $\alpha = V_*/{\bar V}_w$, 
the wind velocity (squared) in the
ambient frame becomes
$$ V'^2_{w} = {\bar V}_w^2 \, 
\biggl[\frac{g_w^2
}{f_w^2
} + 2 \alpha \frac{g_w
}{f_w
}\cos\theta + \alpha^2\biggr].\eqno(123)$$
The wind density is given by
$$ 4 \pi R^2 \rho_w = \frac{{\dot M}_w}{{\bar V}_w} 
\frac{f_w^2
}{g_w
}.\eqno(124)$$
The total kinetic energy flux onto the shell in the ambient frame is
$$ {\dot E}_{\rm therm} =  \frac{{\dot M}_w\,{\bar V}_w^2}{8\pi}  
\int_0^{2\pi} \int_0^\pi \biggl[\frac{g_w^2
}{f_w
} + 2 \alpha g_w
\cos\theta + \alpha^2 f_w
\biggr] \, \sin\theta_*\, d\theta_*\, d\phi_*.\eqno(125)$$
The normalization condition for $f$ permits the last term to be
easily integrated. The second term is $V_*$ times the $z$ component
of the wind momentum flux. In terms of the total wind kinetic energy
flux and momentum flux, we have
$${\dot E}_{therm} = \frac{1}{2}{\dot M}_w (<V_w^2> + V_*^2) +
{\bf P}_w \cdot {\bf V}_*.\eqno(126)$$

We see that a sufficient condition for the total thermalization 
rate to be independent of the wind orientation is the vanishing
of $P_{z,w}$. Because we do not wish to invoke a special alignment
of the wind orientation with the star's direction of motion,
more typically this requires a point-symmetric wind such that
${\bf P}_w = 0$. As a special case, this can be more explicitly 
confirmed for the axisymmetric wind, assuming it is symmetric
with respect to its midplane $\theta_* = \pi/2$.
Because we are considering the total thermalization rate, we may perform
the solid angle integration in starred coordinates. In this case,
we need only the transformation equation for $\cos\theta$ which is given by
eq.(74).
We assume the wind to be axisymmetric and symmetric with respect to 
$\theta_*=\pi/2$, so it has an expansion in $\cos\theta_*$ with only even
powers. This implies that the $2 \alpha \cos\theta$ term in the integral will
give no contribution, by parity, because it yields only a $\sin\phi_*$ term, 
which vanishes in the azimuthal integration  and terms with
odd powers of $\cos\theta_*$, which vanish in the polar angle integration.
The remaining contribution is then
$$ {\dot E}_{\rm therm} =  \frac{{\dot M}_w\,{\bar V}_w}{4}  
\int_0^\pi \biggl[\frac{g_w^2(\theta_*)}{f_w(\theta_*)}
+ \alpha^2 f_w(\theta_*)\biggr] \, \sin\theta_*\, d\theta_*,\eqno(127)$$ 
$$= \frac{1}{2}{\dot M}_w (<V_w^2> + V_*^2).\eqno(128)$$
The total energy thermalization rate is independent
of the orientation of the midplane-symmetric, axisymmetric wind. 
For the case of a non-point-symmetric wind, including 
an axisymmetric wind that doesn't possess mirror symmetry 
with respect to the equator, the total thermalization rate will 
depend upon the stellar orientation 
because of the contribution from the $2\alpha \cos\theta$ term.
This is the case of the example solution given in Section~4.3,
for the wind with quadratic dependence on $\cos\theta_*$.
For that wind, the total vector momentum loss rate is
$$ {\bf P}_w = {{{\dot M}_w {\bar V}_w} \over {3}} c_1 \,{\hat {\bf z}}_*,\eqno(129)$$
where ${\hat {\bf z}}_*$ points in the direction of the wind symmetry axis
$z_*$. The thermalization rate for the bow shock driven by this wind
therefor has a fluctuation depending on orientation of the wind 
symmetry axis with respect to the
stellar velocity of magnitude
$$ \Delta {\dot E}_{therm} = {{{\dot M}_w {\bar V}_w V_*} \over {3}} c_1
\cos\lambda.\eqno(130)$$
If the quadratic term in the wind momentum vanishes ($c_2 = 0$), then
the absolute value of $c_1$  may not exceed unity. The amount this
changes the total thermalization rate then depends upon the parameter
$\alpha = V_*/V_w$.

We must note that 
changing orientation of the point-symmetric wind did not change the total 
thermalization rate, but the spectrum  of shock emission will clearly
be different. For example the peak post-shock temperature
will depend upon the incident normal component of velocity. The sum
of thermalization by shock, relaxation, and mixing is independent of
orientation, but the individual contributions will vary. This highlights
the fact that the properties of bow shocks due to non-axisymmetric
winds may be determined by measuring the bow shock shape, mass surface
density (or column density), kinematics, and radiated energy. Although 
detailed shock calculations are beyond the scope of this work, a substantial
amount of information should be obtainable from these global quantities
and may be sufficient for sources where the data are sparse.

\section{Summary}

I have shown how to solve the problem of non-axisymmetric bow shocks
and wind collisions with a simple formalism that requires 
only algebraic equations. 
Most often one considers constant wind speed for such problems, and
in this case the method leads to exact, analytic (although possibly
implicit) solutions. The availablility of simple analytic solutions 
makes it much easier to model observed sources and derive the properties 
of the driving winds. Among the principal applications of these 
solutions would be to determine the cause of the asymmetry in observed 
bow shocks, whether it be due to the ambient medium or an anisotropic wind, 
or both.

Future improvements are needed to include non-radial and 
accelerating winds and 
shearing motions in the shell, in which case the fluid elements 
are not restricted to a plane. 
Also, non-axisymmetric bow shocks 
due to colliding winds in binary systems, including the orbital motion, 
require a formulation in a non-inertial frame. A forthcoming paper
will describe how to solve the general problem of colliding 
winds from two stars, including both anisotropy and acceleration
in the winds.

\acknowledgments
Portions of this work were done as an NRC Research Associate
at NASA Ames Research Center, and at IPAC with support 
by Long Term Space Astrophysics grant to D.Van Buren. IPAC is
operated  by JPL and the California Institute of Technology
under contract with the National Aeronautics and Space Administration. 
I am greatful for helpful discussions
with J.Anderson and  C.McKee.

{}

\vfil\eject

\figcaption{Thin Shell Model for a Stellar Wind Bow Shock. 
The top panel defines the spherical coordinate system and shows
a thin wedge cut by two planes of constant $\phi$, while the bottom panel
shows the wind, ambient, and tangential flows in the frame of the star.
\label{fig:fig1}}

\figcaption{Solution of eq.(60): Scaled x-coordinate versus that
of a bow shock in a uniform medium.\label{fig:fig2}}

\figcaption{Solution surfaces for $l=3$ (top) and $l=1$ (bottom). Stars mark
the source of the wind. The ambient density increases towards the
left of the page, and the direction of stellar motion is downwards. 
The bow shocks are viewed from an angle of $10^\circ$ by rotating about
the x-axis. Contours of
constant $\theta$ are at every $8^\circ$ up to $120^\circ$, while
contours of constant $\phi$ are at every $10^\circ$. \label{fig:fig3}}

\figcaption{Bow Shock for $c_2 = 3$, a polar wind.
The wind is inclined with respect to the z-direction 
by angles (left column, from top to bottom) 
0,10,30 degrees, and (right column, top to bottom) by 50,70, and 90 degrees.
\label{fig:a}}

\figcaption{Bow Shock for $c_2 = 1.5$, a polar wind.
The wind is inclined by angles 0,10,30,50,70, and 90 degrees
with respect to the z-direction as in Figure~4.\label{fig:b}}

\figcaption{Bow Shock for $c_2 = -1.5$, an equatorial wind.
The wind is inclined by angles 0,10,30,50,70, and 90 degrees
with respect to the z-direction as in Figure~4.\label{fig:e}}

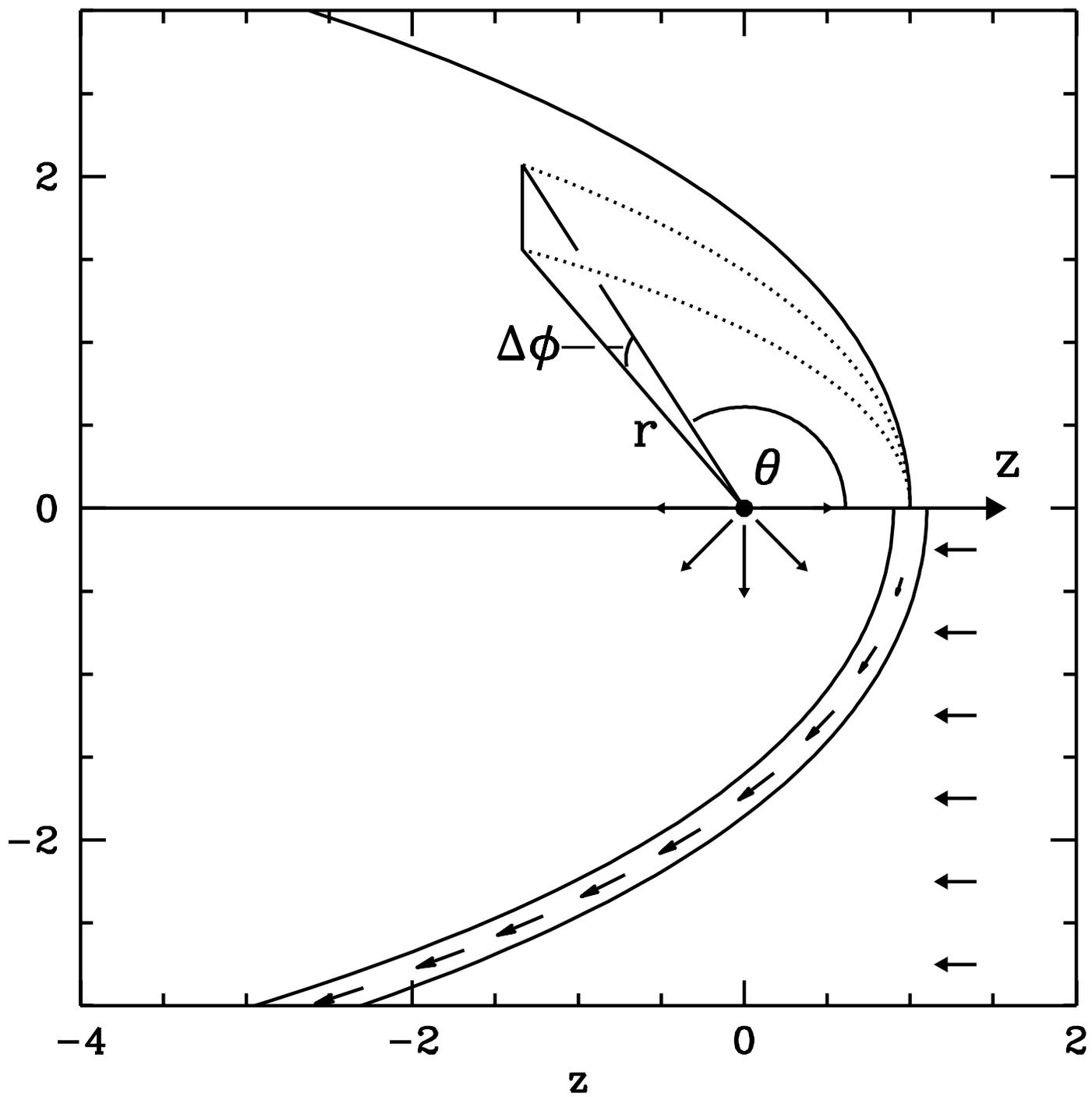
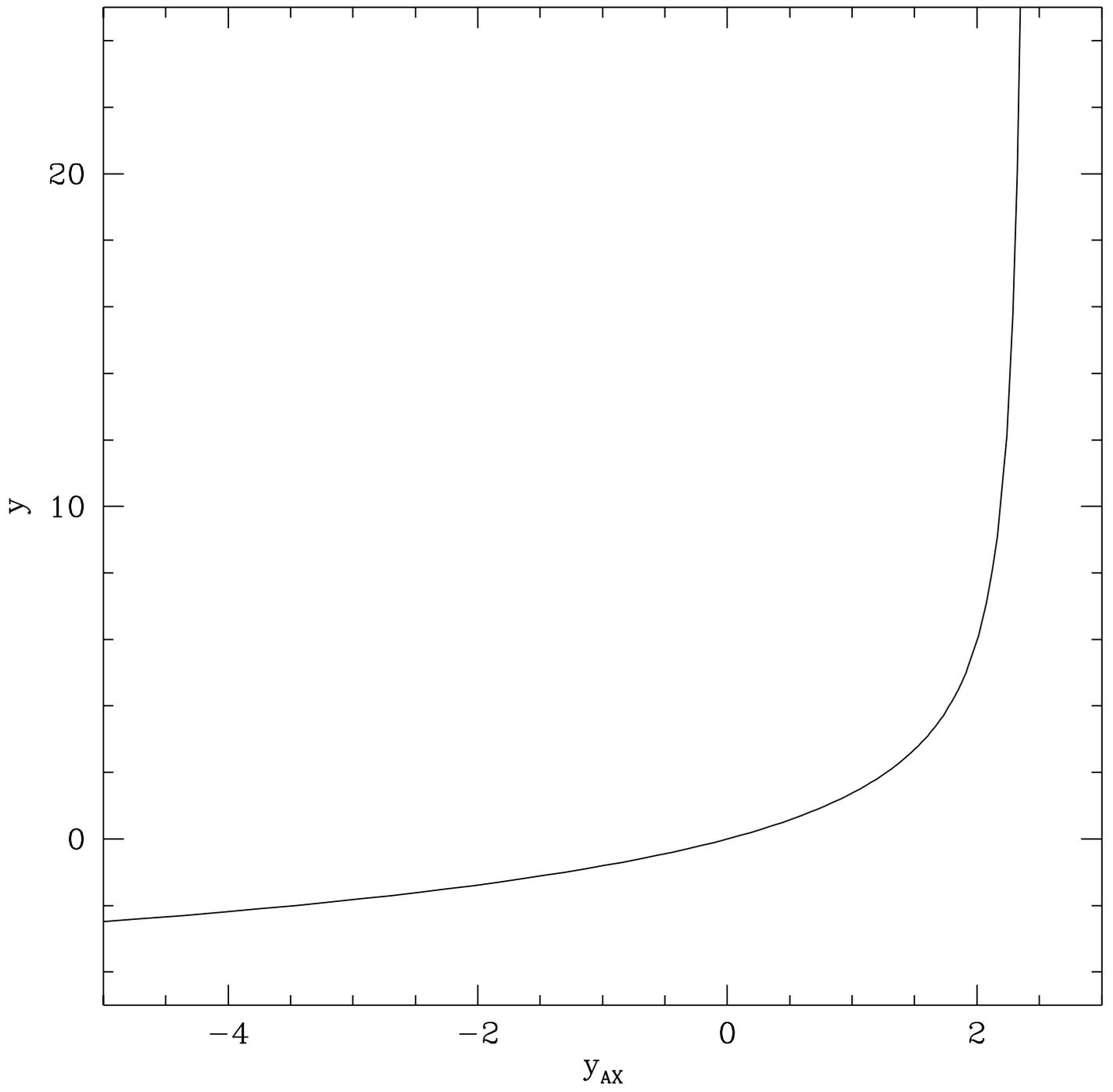
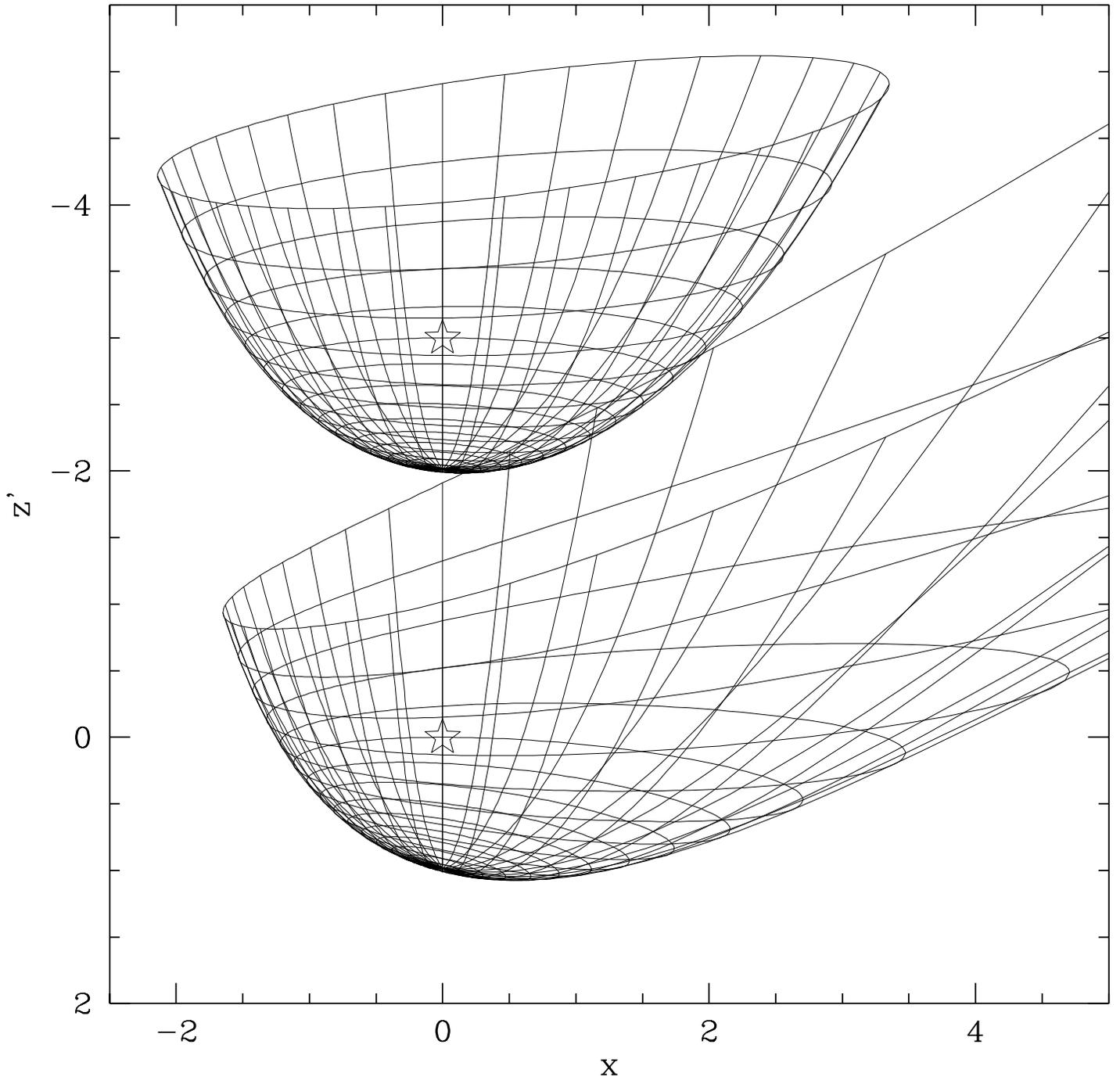
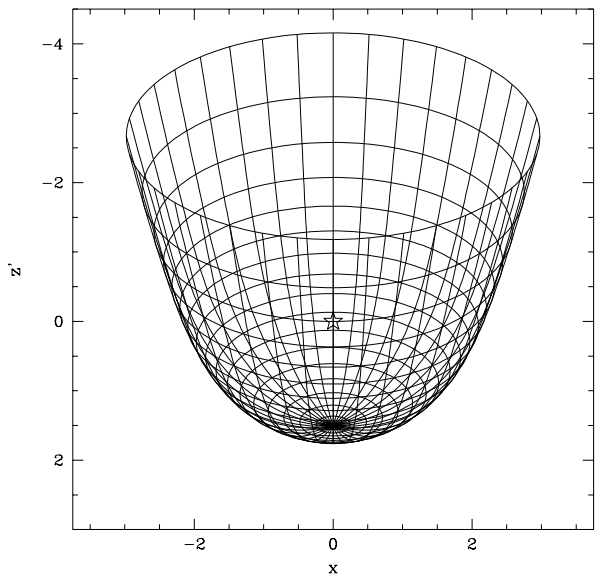
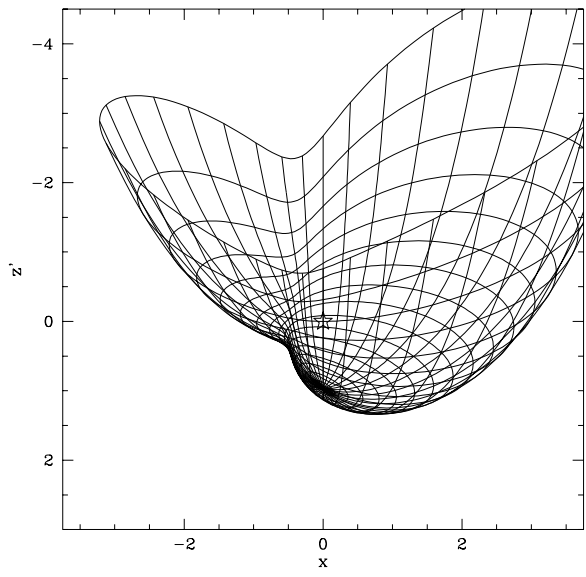
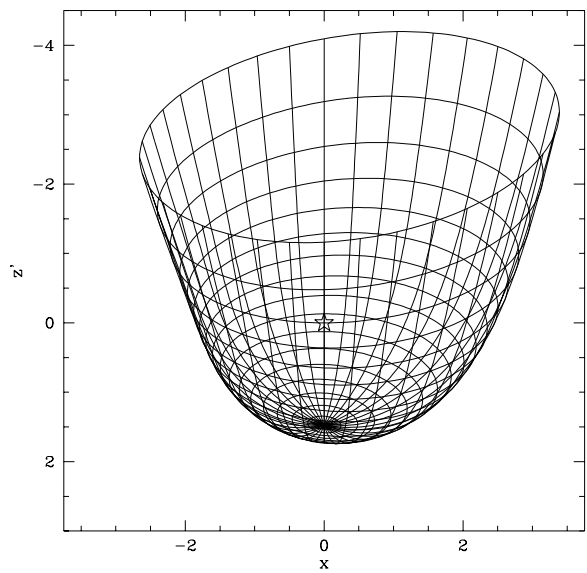
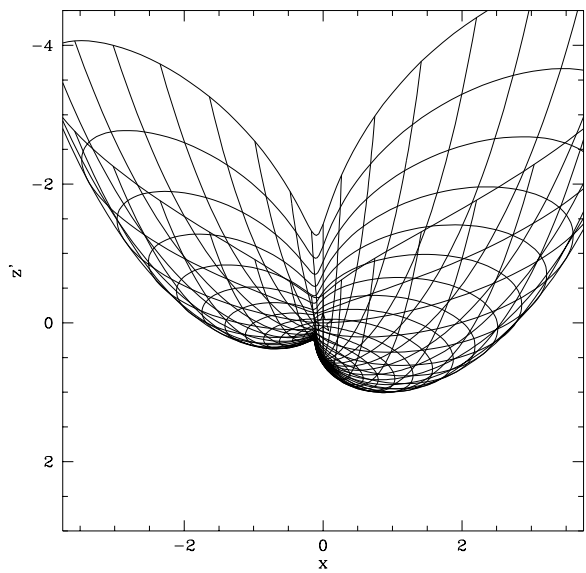
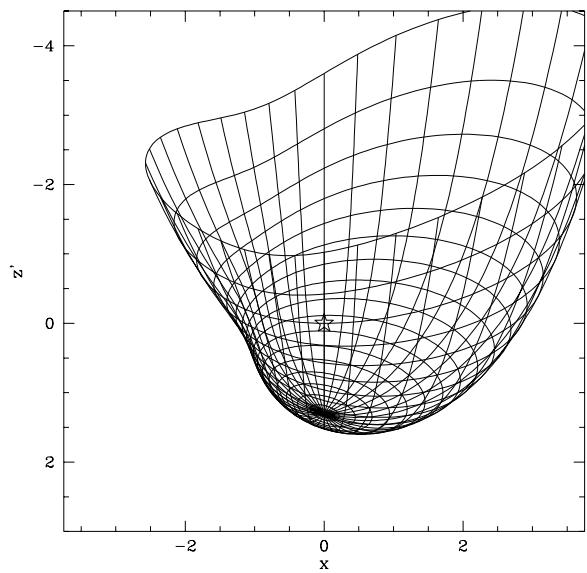
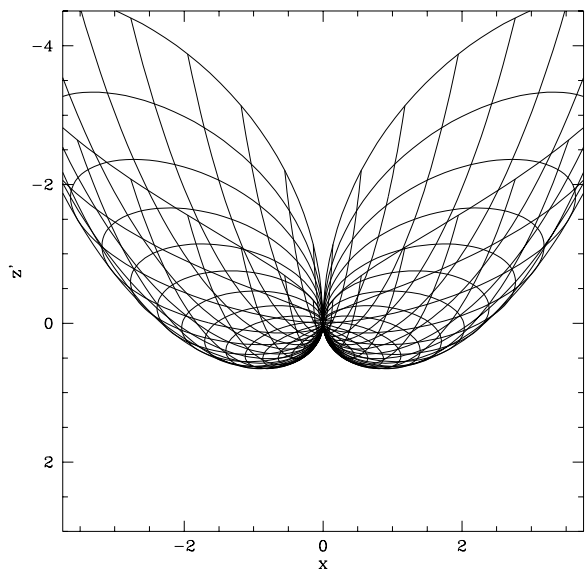
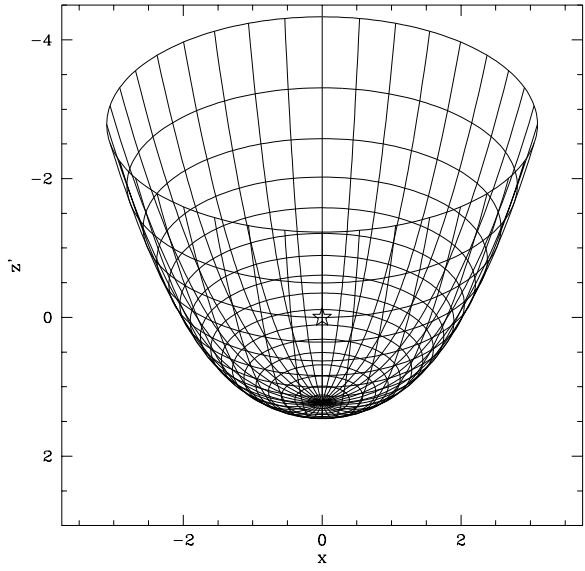
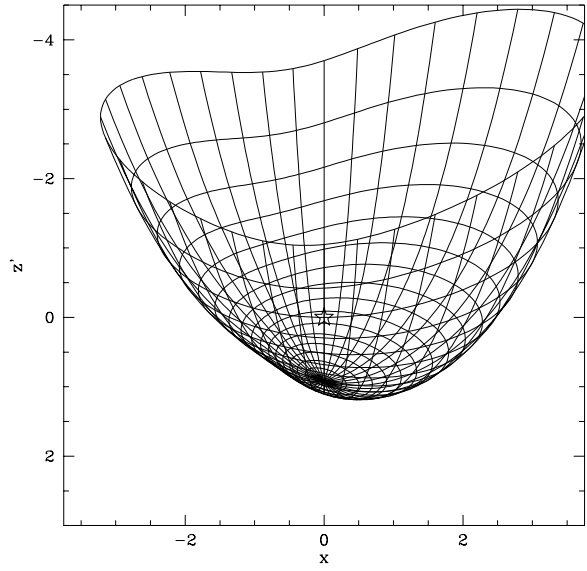
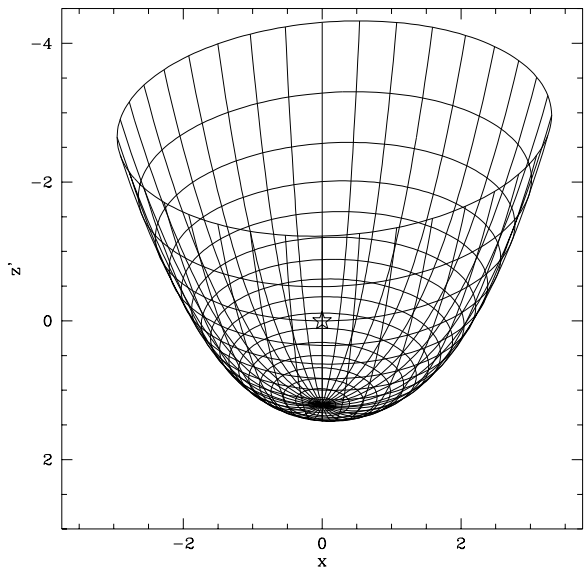
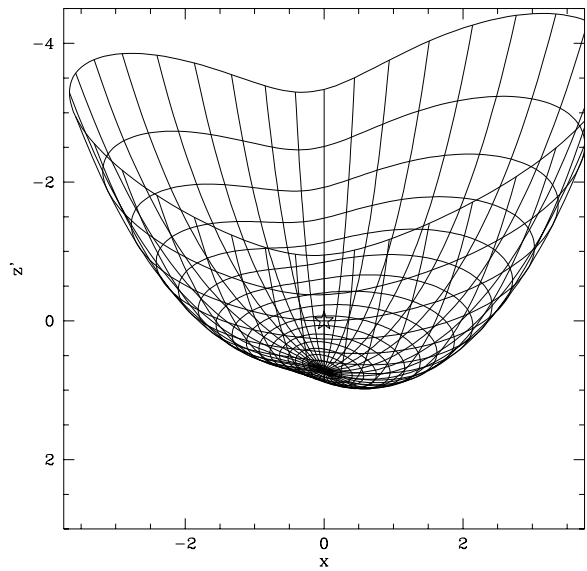
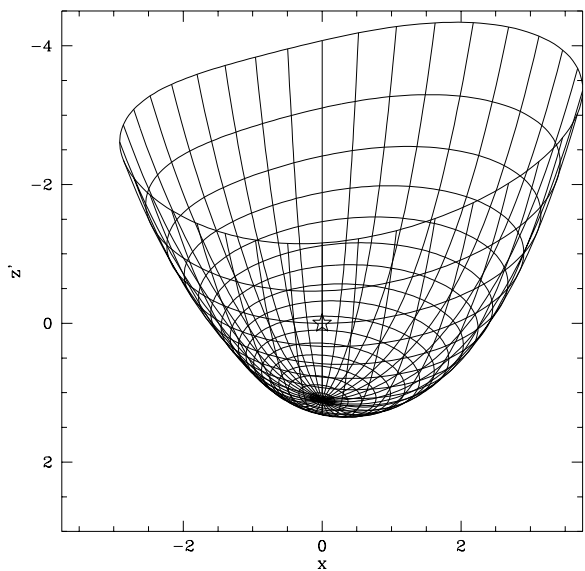
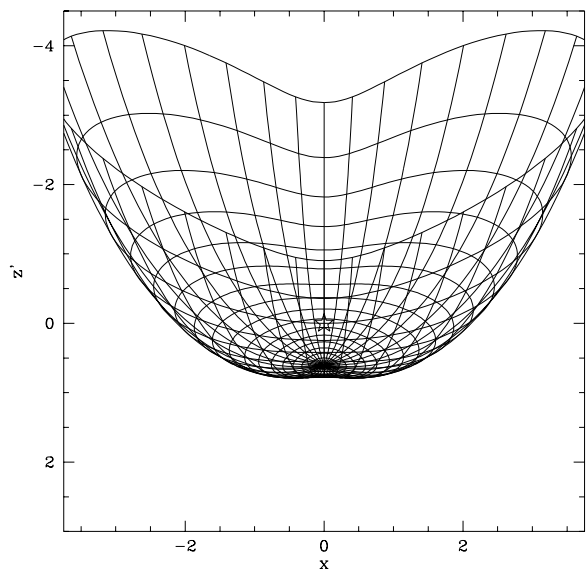
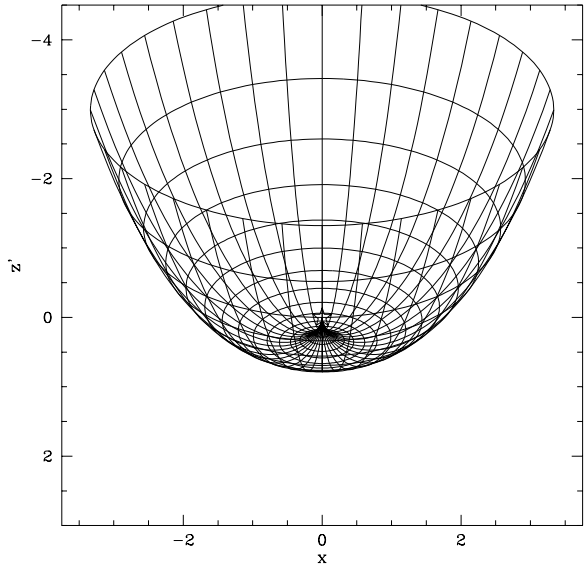
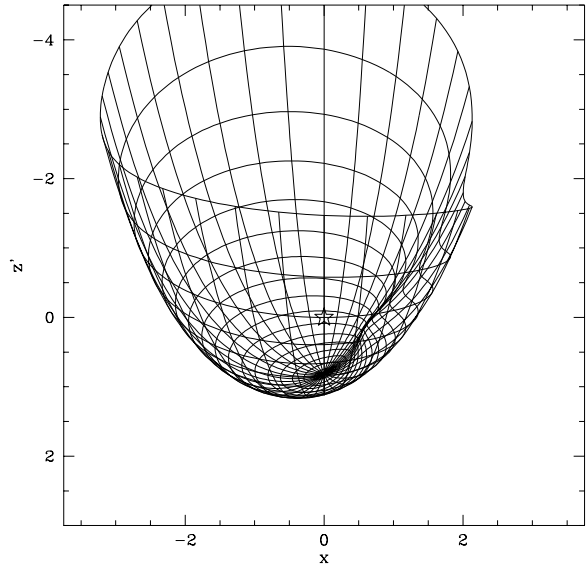
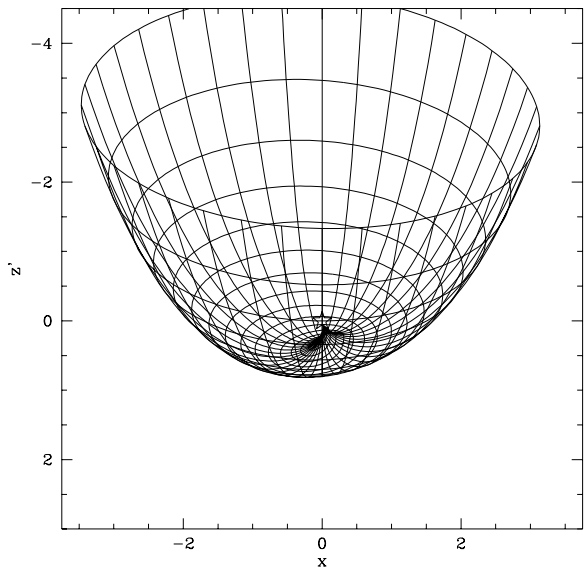
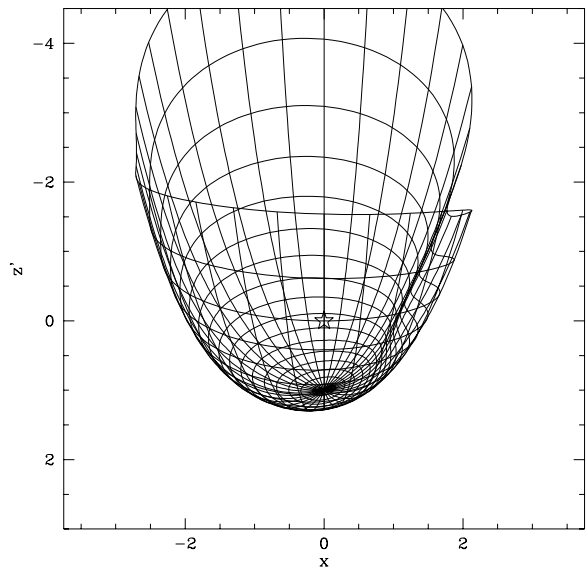
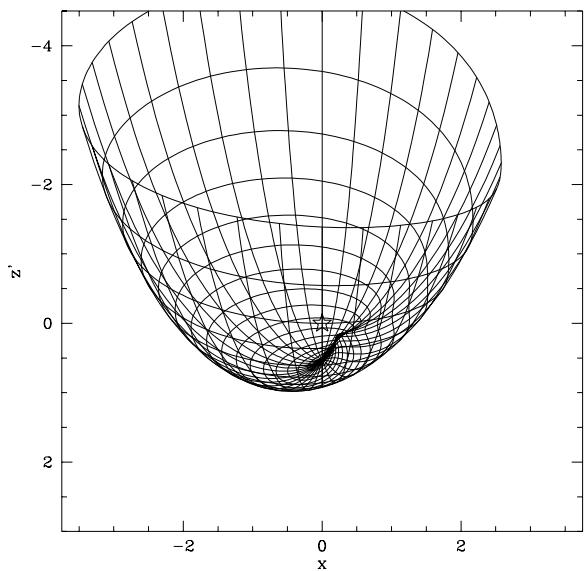
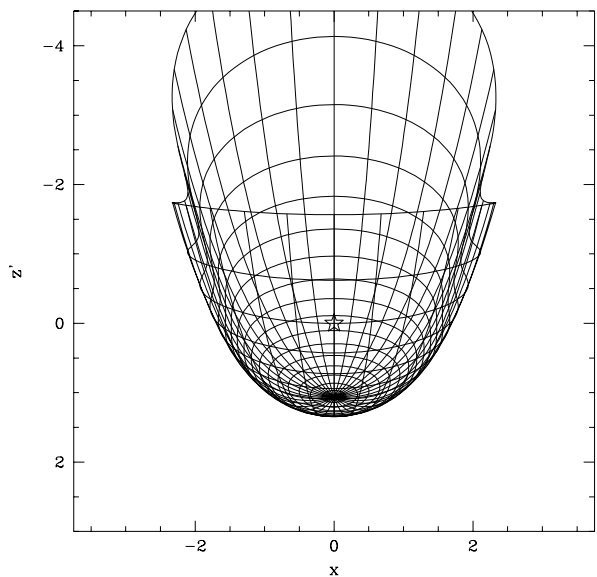
\end{document}